\documentclass{sig-alternate}
\usepackage{color}
\usepackage{enumitem}
\pdfoutput=1
\begin{document}
%
\conferenceinfo{First Int'l Workshop on Open Data}{Nantes, France, 2012}

\title{Publishing Life Science Data as Linked Open Data: the Case Study of miRBase}
%
%
%
%
%

\numberofauthors{5} 
%
\author{
%
%
\alignauthor
Theodore Dalamagas\\
       \affaddr{"Athena" R.C., GR}\\
       \email{dalamag@imis.athena-innovation.gr}
\alignauthor
Nikos Bikakis\\
       \affaddr{"Athena" R.C., GR}\\
       \email{bikakis@imis.athena-innovation.gr}
\alignauthor
George Papastefanatos\\
       \affaddr{"Athena" R.C., GR}\\
       \email{gpapas@imis.athena-innovation.gr}
\and  
\alignauthor
Yannis Stavrakas\\
       \affaddr{"Athena" R.C., GR}\\
       \email{yannis@imis.athena-innovation.gr}
%
\alignauthor
Artemis G. Hatzigeorgiou\\
       \affaddr{"A. Fleming" B.S.R.C., GR }\\
       \email{hatzigeorgiou@fleming.gr}
}

\maketitle
\begin{abstract}
This paper presents our Linked Open Data (LOD) infrastructures for genomic and experimental data related to microRNA biomolecules. Legacy data from two well-known microRNA databases with experimental data and observations, as well as change and version information about microRNA entities, are fused and exported as LOD. Our LOD server assists biologists to explore biological entities and their evolution, and provides a SPARQL endpoint for applications and services to query historical miRNA data and track changes, their causes and effects.
\end{abstract}




\section{Introduction}

The technology advances in scientific hardware (sensors, new-generation sequencers, etc.), together with the explosion of Web 2.0 technologies, have completely changed the way scientists create, disseminate and consume large volumes of information and new content. More and more scientific datasets break the walls of “private” management within their production site, are published, and become available for potential data consumers, i.e., individual users, scientific communities, applications/services. Typical examples include experimental or observational data and scientific models from the life science domain, climate, earth, astronomy, etc.

Linked Data\footnote{http://linkeddata.org/} is a compelling approach for the dissemination and re-use of scientific data, realizing the vision of the so-called Linked Science\footnote{http://linkedscience.org/}. The Linked Data paradigm involves practices to publish, share, and connect data on the Web, and offers a new way of data integration and interoperability. Briefly, Linked Data is about using the Web to create links between data from different sources. The driving force to implement Linked Data spaces is the RDF technology. The basic principles of the Linked Data paradigm is (a) use the RDF data model to publish structured data on the Web, and (b) use RDF links to interlink data from different data sources. The aim of the Linked Data technologies is to give rise to the Web of Data. 

The Web of Data is impelled by the current trend towards an open Web. The open data movement is a significant and emerging force towards this direction. Open science data is open data related to observations and results of scientific activities, which are publicly available for anyone to analyze and reuse.  

However, by just converting legacy scientific data as Linked Open Data (LOD), we do not fully meet the requirements of data re-use. To ensure re-use and allow exploitation and validation of scientific results, several challenges related to scientific data dynamics should be tackled. Scientific data are evolving and diverse data. Users and services (a) should have access not only to up-to-date scientific LOD bases but to any of the previous versions of those bases, and (b) should be able to track the changes among versions, as well as their cause and effects.

In this work, we present our LOD services for life science data, and more specifically, genomic and experimental data related to microRNA biomolecules (see Section 2). Legacy data from two well-known microRNA databases are fused and exported as LOD. The first database (see Section 3)  provides experimental data and observations, while the second one (see Section 4) provides change and version information about microRNA entities.   
Our LOD services provide the following facilities:
\begin{itemize}[topsep=2pt, partopsep=0pt, itemsep=2pt, parsep=0pt]
\item Biologists can explore biological entities, their characteristics, and related experimental data with up-to-date information.
\item Services and applications can retrieve the same (up-to-date) information as above by using our server as SPARQL endpoint. 
\item Biologists can retrieve out-of-date resource descriptions, navigate between previous and next versions of the resources, see the changes involved, their causes and their effects on the resources.
\item Services and applications can retrieve historic information as above by using our server as SPARQL endpoint. 
\end{itemize}
 
The system has been built using the D2R LOD infrustructures\footnote{http://www4.wiwiss.fu-berlin.de/bizer/d2r-server/}. All services are available at \texttt{http://diwis.imis.athena-innovation.gr/mlod}.

\vspace{1ex}\noindent\textbf{Related Work.} Our approach is specially-tailored to the scientific domain of life science data, and more specifically to genomic and experimental data related to microRNA biomolecules.  
Several attempts have been recently made to provide scientific LOD services. W3C has established the Semantic Web Health Care and Life Sciences Interest Group (HCLS)\footnote{http://www.w3.org/blog/hcls/}, aiming to exploit Semantic Web technologies for the management and the representation of biological, medicine and health care data.  The HCLS group works on Linking Open Drug Data (LODD) project which provides linked RDF data exported from several data sources like ClinicialTriasl.gov, DrugBank, DailyMed, etc. Additionally, Bio2RDF\footnote{http://bio2rdf.org/} provides linked RDF data produced from over 30 biological data sources. Some earlier efforts include YeastHub \cite{YEASTHUB}, LinkHub \cite{LINKHUB}, BioDash \cite{BIODASH} and BioGateway\footnote{http://www.semantic-systems-biology.org/biogateway}. Finally, Chem2Bio2RDF \cite{CHEM2BIO2RDF} integrates chemical and biological information. Also, several chemogenomics repositories have been transformed into RDF and linked to Bio2RDF and LODD RDF resources.

In the context of LOD, numerous approaches have been proposed to study the problems of evolution, versioning, and change detection. Particularly, in \cite{UmbrichHHPD10}, the term dataset dynamics is coined, essentially addressing content and interlinking changes in linked data sources. In \cite{Umbrich10}, a comparative study on the approaches and tools for detecting, propagating and describing changes in LOD resources and datasets is provided. In \cite{PopitschH10}, the authors deal with changes in the linkage between datasets and specifically with the problem of broken links. A similar approach is the Silk linking framework \cite{VolzBGK09}, which is used for discovering and maintaining data links between web data sources. Regarding versioning and temporal approaches to LOD, in \cite{SompelSNBSA10} the Memento framework is introduced as a resource versioning mechanism for LOD. Finally, in \cite{Correndo2010} they propose linked timelines, a temporal representation and management for LOD. 

\section{Background}

Biologists used to consider proteins and DNA as movers and shakers in genomics, seeing RNA as nothing more than a messenger to carry information between the two. This has dramatically changed after the discovery, in early 2000s, of the key role played in gene expression by small RNA molecules, called \emph{microRNAs} (miRNAs).  miRNAs can completely silence proteins. They do so by binding themselves to complementary sequences on message RNA (mRNA) transcripts, called \emph{targets}. The knowledge of \emph{miRNA targets} (i.e., which mRNA transcripts are targeted by a miRNA) is important for therapeutic uses. For example, based on such knowledge, biologists can shut off genes by delivering artificial miRNA molecules into cells.

The first miRNA molecules were identified in 1993. Since then, there has been a dramatic increase in the number of miRNAs discovered and registered in \emph{miRBase}\footnote{http://www.mirbase.org/}, a searchable database of published miRNA sequences and annotation. 
However, there is a lack of high-throughput experimental methods for identifying miRNA targets. Thus, \emph{computational methods} to predict targets have become increasingly important, and led to the experimental identification of many miRNA targets.

Our team in IMIS/``Athena'' R.C. and the DNA Intelligent Analysis (DIANA) group of ``Alexander Fleming'' B.S.R.C.\footnote{http://www.fleming.gr/} have developed a set of advanced Web applications to provide access to computationally predicted miRNA targets. 
Since its original launch, DIANA Web app has been one of the most widely used service for miRNA analysis. It includes the following two core services.

\noindent\textbf{microT}\footnote{http://diana.cslab.ece.ntua.gr/DianaTools\\/index.php?r=microtv4}. The service provides target prediction data for 1884 miRNAs and more than six million predicted target genes, organized in a relational database.  
Besides the target prediction experimental results, we provide miRNAs and genes functional analysis that goes beyond simple biological pathways, like, for example,  relation of miRNAs to functional features, and diseases and medical descriptors.
All retrieved miRNAs are associated to diseases, using textual information from PubMed\footnote{http://www.ncbi.nlm.nih.gov/pubmed/}, a well-known digital library for biomedical literature. 

\noindent\textbf{mirGen}\footnote{http://diana.cslab.ece.ntua.gr/?sec=databases}. The service provides information ab\-out tra\-nscri\-pts, and their transcription factors (TF) that correspond to miRNAs. A transcription factor is a protein that binds to specific DNA sequences, thereby controlling the flow of genetic information from DNA to mRNA. MirGen database stores information about 811 human genes, 1270 human miRNAs, 386 mouse genes and 1012 mouse miRNAs, organized in a relational database.

\section{Database Overview}

Next, we present an overview of the miRNA database maintained by our team in IMIS/“Athena” R.C. and the DIANA group of ``A. Fleming'' B.S.R.C., storing info about computationally predicted miRNA targets produced by the target prediction algorithm proposed by DIANA group\cite{maragkakis2009}.

To better understand the miRNA domain and the DB schema design, we next clarify some issues.
Since the term ``miRNA'' is nowadays used in a wide scope, it is common to distinguish between
\texttt{hairpin miRNAs} and \texttt{mature miRNAs}, or just \texttt{hairpins} and \texttt{matures} from now on. The former signifies the genomic location of the latter. A hairpin is actually processed into several matures. Matures bind themselves to transcripts and prevent the creation of functional ribosomes (and, thus,  prohibit protein construction). A transcript is a stretch of DNA transcribed into an RNA molecule (messenger RNA, ribosomal RNA, transfer RNA, etc). 

The miRNA database has some core tables to store the key entities of the miRNA domain (hairpins, matures, transcripts and protein-encoding genes) and model their relationships (see Table \ref{tab:db} for a part of miRNA database schema). 
\begin{table}
{\center
\scriptsize{
\begin{tabular}{ | l | p{4.5cm} |}
\hline
\textbf{Core tables} & \textbf{Column Description}  \\ \hline
Hairpins & id (\texttt{mima\_id}), name, sequence, species, gene location info, etc.\\ \hline
Matures & id (\texttt{mimat}), name, sequence, species. \\ \hline
Transcripts & tid, id given from ensembl.org (\texttt{enstid}), species, DNA strand, gene location info, etc. \\ \hline
ProteinGenes & id given from ensembl.org (\texttt{ensgid}), name, description.\\ \hline                  
Keggs & id given from genome.jp (\texttt{kegg\_id}), name. \\ \hline                           
Tissues & name, species. \\ \hline                           
\textbf{Join Tables} & \textbf{Column Description}  \\ \hline
MatureHairpinConn & It relates matures and hairpins.\\ \hline       
MicroT5Interactions  & It contains all the experimentally verified gene-mature interactions (bindings). \\ \hline         
ProteinGeneKeggConn & It relates genes to kegg pathways. \\ \hline 
MatureTissueConn & It relates matures to tissues. \\ \hline 
\end{tabular}
\caption{Part of miRNA database schema.}
\label{tab:db}
}
}
\end{table}
There are also tables storing info about \texttt{Kegg pathways}\footnote{http://www.genome.jp/kegg/pathway.html} and \texttt{tissues}. Kegg pathways is a collection of manually drawn pathway maps, with textual descriptions, representing biologists’ knowledge on molecular interaction and reaction networks.

\section{Change and version management}

The miRBase database is a searchable
database of published miRNA sequences and annotation. 
The miRBase database maintains info for $18443$ hairpins and $49670$ matures.  
Each entry in miRBase
represents a predicted hairpin miRNA with information on the location and sequence of
the corresponding mature miRNA sequence. 
Hairpins, mature miRNAs and their relationship between them change in time. miRBase maintains a list of files that record successive versions along with the changes between them. A short description for each file follows. 

\begin{figure}[!ht] \center
   \includegraphics[width=8.5cm]{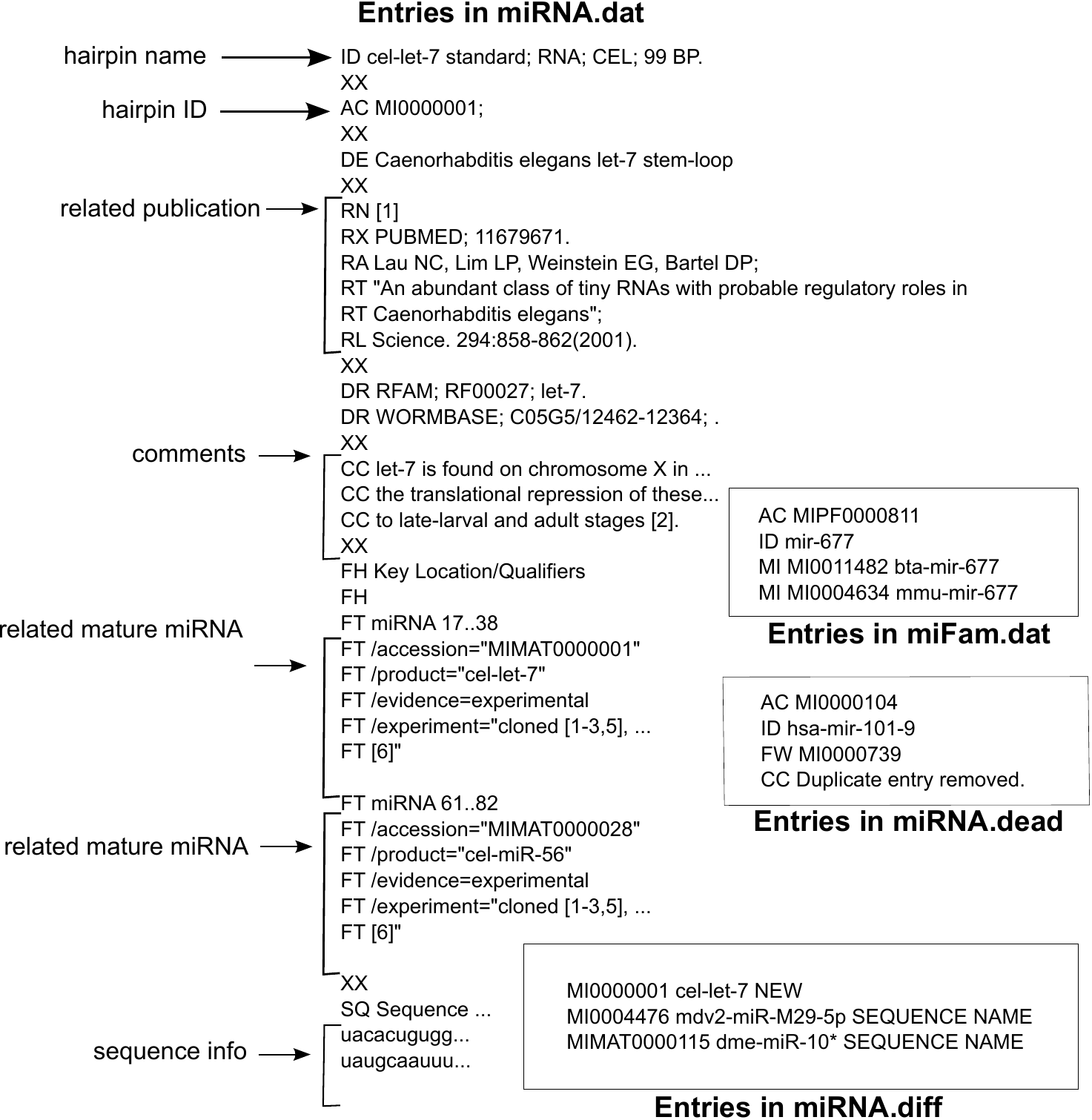}          
    \caption{File examples of tracking miRNA changes.}
    \label{fig:miRNAdata-example-short}
\end{figure}

\begin{itemize}[topsep=2pt, partopsep=0pt, itemsep=2pt, parsep=0pt]
\item\textbf{miRNA.dat} It maintains info related to all known
hairpins (like ID, name, related matures, related publications, sequence,
etc.) at the time of each version. 
Every new version of miRNA.dat contributes to the previous one with all the newly discovered miRNAs, omitting the deleted ones. 
Example entries of miRNA.dat are shown in Figure \ref{fig:miRNAdata-example-short}, where info about the hairpin with name \texttt{cel-let-7} and id (i.e., key) \texttt{MI0000001} is presented.
\item\textbf{miRNA.diff} It tracks change operations on hairpins
and matures. Each version of miRNA.diff refers to a certain time period
and tracks changes only for that period. 

Example entries of miRNA.diff are shown in Figure \ref{fig:miRNAdata-example-short}.
For instance, \texttt{MI0000001 cel-let-7 NEW} means that the hairpin with ID
MI0000001 and name cel-let-7 is created. Also, \texttt{MI0004476 mdv2-miR-M29-5p
SEQUENCE NAME} means that the hairpin with ID MI0004476 has changed its name (to
mdv2-miR-M29-5p) and its sequence. Note that to find the old name and the old
sequence, we should refer to the older version of the miRNA.dat file, where
hairpin names and info about sequences are available. Similarly, 
\texttt{MIMAT0000115 dme-miR-10* SEQUENCE NAME} means that the mature
with ID MIMAT0000115 has changed its name (to dme-miR-10*) and its sequence.
Note that IDs starting with ``MIMA'' refer to matures.
\item\textbf{miRNA.dead} It keeps all deleted hairpins at the time
of a version. It is maintained incrementally. Deletion means either getting rid of a
hairpin (e.g., incorrectly characterized in previous versions) or replacing a hairpin with another one. For the latter case, links to existing
hairpins are provided. Contrary to deleted hairpins, deleted mature miRNAs are not stored in miRNA.dead file. 

Example entries of miRNA.dead are shown in Figure
\ref{fig:miRNAdata-example-short}. For instance, the hairpin with ID
hsa-mir-101-9 and NAME MI0000104 has been deleted. The reason is that it was a
duplicate entry (see the comment in CC field). There is a hairpin (MI0000739),
though, that replaces the deleted one (see the FW field).
\item\textbf{miFam.dat} It stores info about hairpin families
at the time of a version. Hairpins that produce similar mature miRNAs belong to
the same family. It is maintained incrementally.
Example entries of miFam.dat are shown in Figure \ref{fig:miRNAdata-example-short}.
For instance, hairpins with IDs MI0011482 (NAME bta-mir-677) and MI0004634
(NAME mmu-mir-677) belong to the same family with is mir-677. 
\end{itemize}

We have examined all files and recorded the following types of changes for hairpins:
(a) NEW: a new hairpin is created, (b) NAME: a hairpin changes its name, (c) SEQUENCE (SEQ): a hairpin changes its sequence, 
(d) NAME\-/SEQUENCE (NS): a hairpin changes both its name and sequence at the same time, (e) FORWARD (FW): a hairpin is deleted, but miRBase give a link to another hairpin for replacement, and (f) DELETE (DEL): a hairpin is deleted (no replacement). 

Similarly, we have identified the following type of changes for matures:
(a) NEW: a new mature is created, (b) NAME: a mature changes its name, (c) SEQUENCE (SEQ): a mature changes its sequence, 
(d) NAME/SEQUENCE (NS): a mature changes both its name and sequence at the same time, (e) ADD PARENT HAIRPIN (APH): a new hairpin is added to the list of hairpins that produces a mature, (f) REMOVE PARENT HAIRPIN (RPH): a hairpin is removed from the list of hairpins that produces a mature, and (g) DELETE (DEL): a mature is deleted.

To manage change and version info, we maintain two history tables: HairpinsHistory and MaturesHistory. Tables \ref{tab:hversiondb} and \ref{tab:mversiondb} show how change and version info is maintained in history tables. 
For each hairpin change, HairpinsHistory keeps a record with, among others, the hairpin id, the type of change, the version number where the change occurred, and the version number where the next change occurs. 
The hairpin with id ..1364 is first created in version $13$. In version $16$, it changes name from \texttt{dre-mir-10b} to \texttt{dre-mir-10b-1}. No other change has occurred till version $18$, where a change in its sequence has occurred. Another sequence change has occurred in version $20$. 
Similarly, the mature with id ..9477 is first created in version $28$, getting the name \texttt{bfl-miR-79}, and having the parent hairpin ..021. In version $30$, it changes name (to bfl-miR-9-3p) and sequence.  
\begin{table}[!ht]
{\center
\scriptsize{
\begin{tabular}{ | p{0.8cm} | c | c  | c | p{1cm} | p{1cm} |}
\hline
\textbf{mima\-id} & \textbf{change} & \textbf{name} & \textbf{seq} & \textbf{first\_app\-earance} & \textbf{last\_app\-earance} \\ \hline
..1364 & NEW & dre-mir-10b & ..X.. & 13 & 15 \\ \hline
..1364 & NAME & dre-mir-10b-1 & ..X.. & 16 & 17 \\ \hline
..1364 & SEQ & dre-mir-10b-1 & ..Y.. & 18 & 19 \\ \hline
..1364 & SEQ & dre-mir-10b-1 & ..Z.. & 20 & 32 \\ \hline
\end{tabular}
\caption{Table HairpinsHistory: record samples.}
\label{tab:hversiondb}
}
}
{\center
\scriptsize{
\begin{tabular}{ | p{0.8cm} | p{0.8cm} |  c  |  c  | p{0.6cm}  | p{1cm} | p{1cm} |}
\hline
\textbf{mimat} & \textbf{change} & \textbf{name} & \textbf{seq} & \textbf{par. hair\-pin} & \textbf{first\_app\-earance} & \textbf{last\_app\-earance} \\ \hline
..9477 & NEW & bfl-miR-79 & .X. & .. & 28 & .. \\ \hline
..9477 & APH & bfl-miR-79 & .X. & ..021 & 28 & 29 \\ \hline
..9477 & NS & bfl-miR-9-3p & .Y. & .. & 30 & 32 \\ \hline
\end{tabular}
\caption{Table MaturesHistory: record samples.}
\label{tab:mversiondb}
}
}
\end{table}

\vspace{-2ex}
\section{Publishing linked open miRNA data}

\subsection{LOD technology adopted}

To publish miRNA and miRBase databases as LOD, we adopted the ``virtual RDF'' approach: accessing a non-RDF database using an RDF view. Such an approach enables the access  of non-RDF, legacy databases without having to replicate the whole database into RDF. The D2R server \cite{bizer06} is a popular tool that follows the ``virtual RDF'' approach for publishing the content of relational databases on the Semantic Web. Database content is mapped to RDF using the D2RQ declarative mapping language that captures mappings between database schemas and RDFS/OWL schemas. 

A D2RQ mapping specifies how RDF resources are identified and how RDF property values are generated from database content. Mappings in D2RQ are declared based on \emph{ClassMaps} and \emph{PropertyBridges}. A ClassMap maps a set of database records  to an RDF class of resources. 
Resources are assigned URIs using URI patterns. The pattern \texttt{hairpins/$@@$diana\_hairpins.mima\_id$@@$}, for instance,
produces a relative URI like \texttt{hairpins/MI0000005} by inserting a value from the column \texttt{mima\_id} of table \texttt{hairpins} of miRNA database into the pattern. The D2R Server turns relative URIs into absolute URIs by expanding them with the server’s base URI. If a database already contains URIs for identifying database content, then these external URIs can be used instead of pattern-generated URIs. The following ClassMap definition creates the class of hairpin resources, and assigns them URIs using their ids from the miRNA database:

{\scriptsize 
\begin{verbatim}
map:Hairpins a d2rq:ClassMap;
     d2rq:dataStorage map:database;
     d2rq:uriPattern "hairpins/@@diana_hairpins.mima_id@@";        
     d2rq:class diana:Hairpin;
     d2rq:classDefinitionLabel "Hairpin";
\end{verbatim}        
}

Each ClassMap has a set of PropertyBridges which specify how the properties of an RDF instance are created. Property values can be literals, URIs or blank nodes, and can be created directly from database values or by employing patterns. The following PropertyBridge definition creates the property \texttt{diana:name}. Values for that property are created from the \texttt{name} column of table \texttt{diana\_haipins}:

{\scriptsize 
\begin{verbatim}
map:diana_hairpins_name a d2rq:PropertyBridge;
     d2rq:belongsToClassMap map:Hairpins; 
     d2rq:property diana:name;
     d2rq:propertyDefinitionLabel "Hairpins name";
     d2rq:column "diana_hairpins.name";
\end{verbatim}
}

Note that D2R provides flexible mappings of complex relational structures, allowing  SQL statements directly in the mapping rules. The resulting record sets are grouped afterwards and the data is mapped to the created instances.

We used D2R as a full-fledge Linked Open Data server. The size of the LOD base is around 100Million triples.

\subsection{LOD publishing}

\begin{figure*}[!ht] \center
\includegraphics[width=16cm]{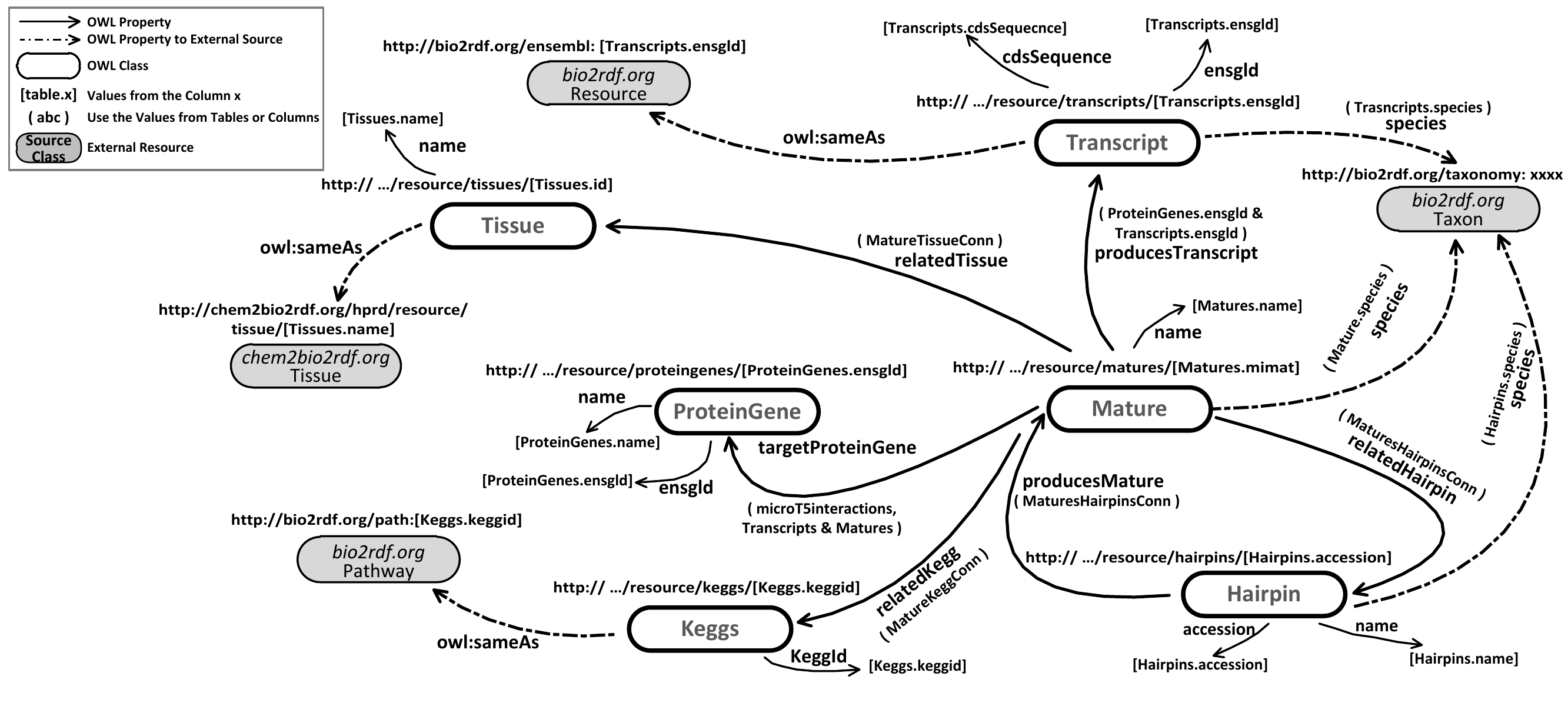}  
    \caption{Publishing miRNA data as LOD data: RDFS to database mappings (up-to-date data).}
    \label{fig:mappings}
%
\includegraphics[width=14.5cm]{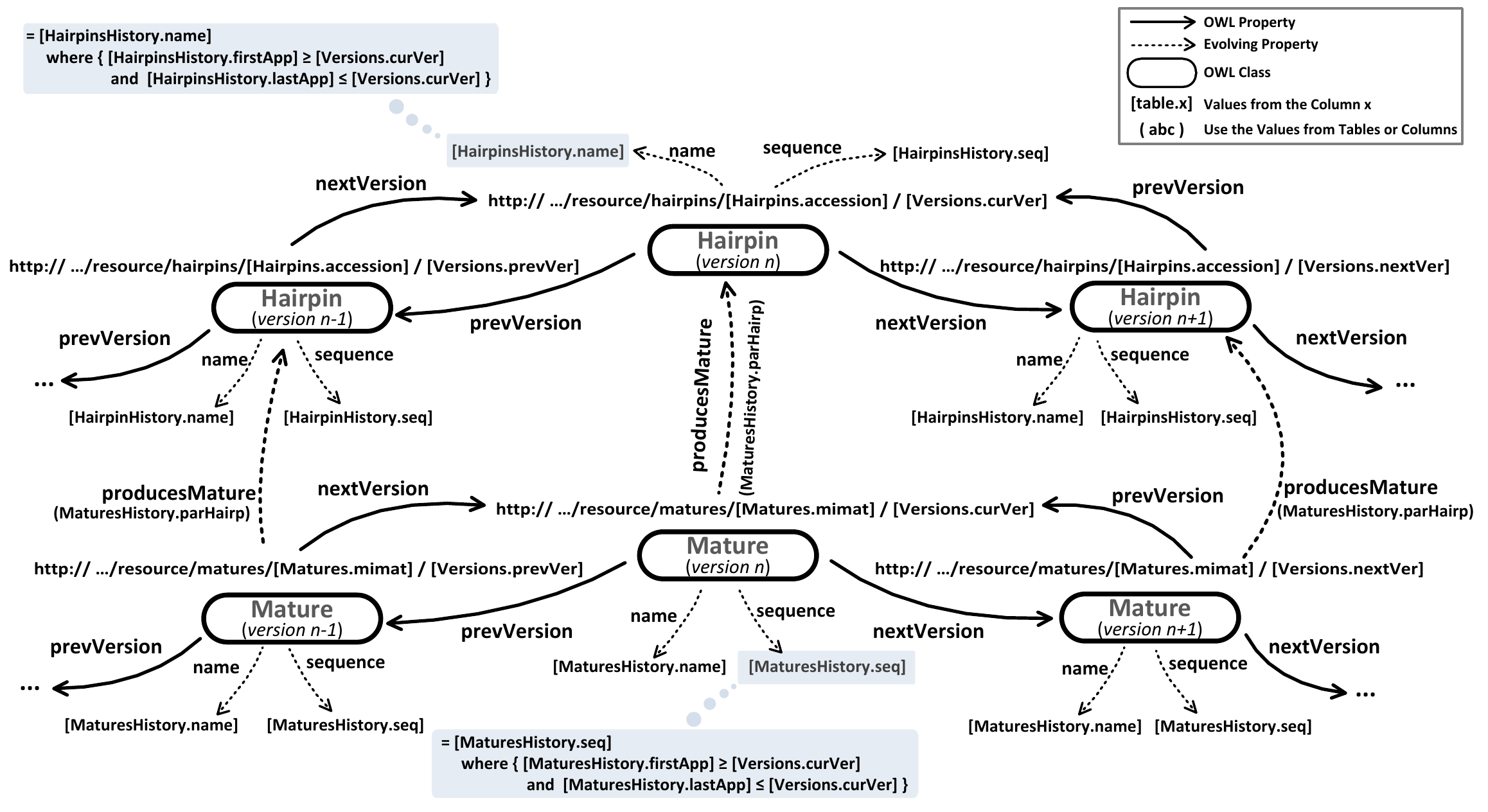}  
\caption{Publishing miRNA data as LOD data: RDFS to database mappings (historic data).}
    \label{fig:mappings-v}
\end{figure*}

The miRNA LOD schema has been designed around four core classes: \texttt{Hairpin}, \texttt{Mature}, \texttt{ProteinGene} and \texttt{Transcript} (defined as ClassMap entities in D2R - see previous subsection). Figure \ref{fig:mappings} shows an overview of the schema adopted and part of the mappings used. Consider, for example, the class \texttt{Mature}. Resources of that class are assigned URIs of the form \texttt{http://.../resource/matures/[Matures.mimat]}, where \texttt{Matures.mimat} gets values from column \texttt{mimat} of Table \texttt{Matures}. Some of the class properties are: \texttt{name}, \texttt{species}, \texttt{relatedKegg},  \texttt{targetsProteinGene} (defined as PropertyBridge entities in D2R - see previous subsection). 

Consider also the property \texttt{targetsProteinGene} that relates matures with genes (targets). Note that \texttt{ProteinGene} resources are assigned URIs of the form \texttt{http://.../re\-sou\-rce/proteingenes/[ProteinGenes.ensgid]}, where \texttt{Pro\-te\-inGenes.ensgid} gets values from column \texttt{ensgid} of Table \texttt{ProteinGenes}. For a given \texttt{Mature} URI, to calculate the URIs of related \texttt{ProteinGene} resources, the mapping definition should include the following join: \\
\texttt{Matures.mimat=MicroT5Interactions.mimat AND \\ 
MicroT5Interactions.tid=Transcripts.tid AND \\
Transcripts.enstid=ProteinGenes.enstid.}

To link our LOD to the LOD cloud, we provide \texttt{owl:sameAs} links to appropriate biological LOD infrastructures. See, for example, the BIO2RDF\footnote{http://bio2rdf.org} data source that provide RDF descriptions for transcripts, tissues, keggs, and species.

\subsubsection{Change and version management}

One of the major research problems in LOD publishing is how to deal with linked data that changes over time. While handling changes for information resources is rather straightforward, handling changes for non-information resources is a challenging issue. Key requirements for dealing with changes in miRNA LOD are the following:
\begin{itemize}[topsep=2pt, partopsep=0pt, itemsep=2pt, parsep=0pt]
\item Biologists that care only about the current state of data should be able to browse or query the miRNA LOD base easily to get up-to-date data. Also, up-to-date data should be easily retrieved using SPARQL.
\item Biologists should be able to query historic miRNA data, and navigate through versions. Also, miRNA changes should be treated as first-class citizens so that one can form SPARQL queries that involve change resources, and trace those changes and their effects.
\end{itemize}

\begin{figure*}[!ht] \center
\includegraphics[width=14cm]{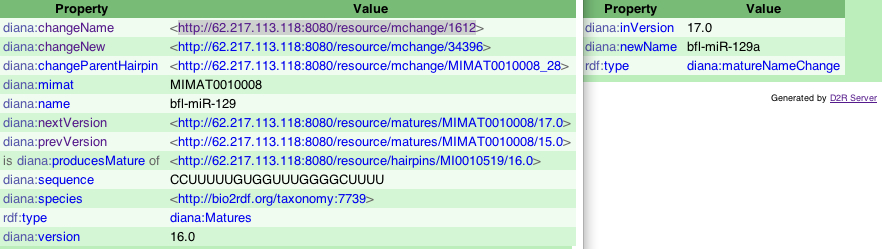}      
\caption{Resource description of mature MIMAT0010008 at version 16.0}
    \label{fig:d2r-sample}
\end{figure*}

\noindent\textbf{Browsing and querying up-to-date miRNA data.} 
Using the D2R browsing facilities, biologists can navigate through the miRNA LOD base, exploring hairpin, mature, gene or transcript resources and their descriptions. All data provided refer to the current version of miRNA database. Also, any resource URI refers to the current version of that resource. This is ensured because all triples involving resources from \texttt{Hairpin}, \texttt{Mature}, \texttt{ProteinGene} and \texttt{Transcript} classes are populated from the core and join tables of Table \ref{tab:db} that are up-to-dated.   
Using the D2R SPARQL end-point facilities, biologists can pose SPARQL queries to the miRNA LOD. Whenever a resource URI is used in a query, it refers to the current version of that resource. To get up-to-date results, a property should be used to avoid the retrieval of out-of-date triples. For example, the following SPARQL query retrieves 10 hairpins, and their sequences, that are located in chromosome X from the current version of miRNA LOD: 

\vspace{-1ex}
{\scriptsize 
\begin{verbatim}
SELECT ?h ?s WHERE {
  ?h rdf:type diana:Hairpin.
  ?h diana:sequence ?s.
  ?h diana:chromosome "X".
  ?h diana:label "now". } LIMIT 10
\end{verbatim}
}

\noindent\textbf{Browsing and querying historic miRNA data.}
Out-of-date resource descriptions are retrieved using the following pattern for URIs: \texttt{URI/\{ve\-rsion number\}}. For example, the URI \texttt{http://.../resour\-ce\-/hairpins/MI0000044/8.0} gets the RDF description of hairpin \texttt{MI0000044} in version $8.0$ of miRBase. To pose the previous SPARQL query on that version of miRBase, one should replace \texttt{?h diana:label "now".} with \texttt{?h diana:version "8.0".}  Note that we provide properties (\texttt{diana:nextVersion}, \texttt{diana:prevVersion}) to move to the next and the previous version of a resource description.

To be able to provide the property values and URIs which are valid at a certain version, we exploit the version information present in the history tables HairpinsHistory and MaturesHistory (see Tables \ref{tab:hversiondb} and \ref{tab:mversiondb}). Figure \ref{fig:mappings-v} shows an overview of the schema adopted and part of the mappings used to manage changes and versions. For example, given a current version \texttt{curVer}, to retrieve the valid value for the \texttt{name} property of a hairpin, we should define a conditional mapping to focus the retrieval on values that remain unchanged for a time period that starts before \texttt{curVer} and ends after \texttt{curVer} (similarly for, e.g., mature names).

Each hairpin or mature resource description includes properties that capture the changes which those resources are affected by. For each change, we track its effect and its cause. Figure \ref{fig:d2r-sample} shows the description of mature MIMAT0010008 at version $16$.  The following SPARQL query retrieves $10$ hairpins that where deleted or replaced in version in version $1.3.$ of miRBase, and the URIs of the change operations: 

\vspace{-1ex}
{\scriptsize 
\begin{verbatim}
SELECT ?h ?d ?c WHERE {
 ?h rdf:type diana:Hairpin.
 {{?h diana:changeDelete ?d.} UNION {?h diana:changeForward ?c.}}
 ?h diana:version "1.3". } LIMIT 10
\end{verbatim}
}
\vspace{-2ex}

\vspace{-1ex}
\section{Conclusion and Further Work}
 
 In this work we presented a case study of publishing genomic and experimental data related to microRNA bio\-mo\-le\-cu\-les as Linked Open Data. Legacy data from two well-known microRNA databases with experimental data and observations, as well as change and version information about microRNA entities, are fused and exported as LOD. The miRNA LOD server assists biologists to explore biological entities, and navigate between previous/next versions of the resources, and also provides a SPARQL endpoint for applications to query historical miRNA data and track changes.
As future work, we plan to expand the LOD set with resources available from gene databanks, and also to implement materialized approaches (i.e., using a native RDF store).

\bibliographystyle{abbrv}
\bibliography{od}  

\end{document}